\documentclass[twocolumn,showpacs,preprintnumbers,amsmath,amssymb]{revtex4}
\usepackage{graphicx}
\usepackage{dcolumn}
\usepackage{bm}
\begin{document}
\title{Lattice Dynamics and Electron-Phonon Interaction in (3,3) Carbon Nanotubes}
\author{K.-P. Bohnen$^1$, R. Heid$^1$, H.J. Liu$^2$, and C.T. Chan$^2$}
\affiliation{$^{1}$Forschungszentrum Karlsruhe, Institut f\"ur
Festk\"orperphysik, P.O.B. 3640, D-76021 Karlsruhe, Germany}
\affiliation{$^{2}$Department of Physics, University of Science
and Technology, Clear Water Bay, Kowloon, Hong Kong, China}

\date{\today}
\begin{abstract}
We present a detailed study of the lattice dynamics and
electron-phonon coupling for a (3,3) carbon nanotube which belongs
to the class of small diameter based nanotubes which have recently
been claimed to be superconducting. 
We treat the electronic and phononic
degrees of freedom completely by modern ab-initio methods without
involving approximations beyond the local density approximation.
Using density functional perturbation theory we find a mean-field
Peierls transition temperature of $\approx$240\,K which is an order of
magnitude larger than the calculated superconducting transition temperature.
Thus in (3,3) tubes the Peierls transition might compete with
superconductivity.
The Peierls instability is
related to the special 2k$_F$ nesting feature of the Fermi
surface. Due to the special topology of the (n,n) tubes also a q=0
coupling between the two bands crossing the Fermi energy at k$_F$
is possible which leads to a phonon softening at the
$\Gamma$ point.

\end{abstract}
\maketitle

During the past 10 years carbon nanotubes have gained a lot of
attention \cite{Saito}. This is due to their potential for
applications (e.g. for molecular electronics) as well as due to
the fact that they allow to study electronic systems in one
dimension. Thus they offer the opportunity to investigate effects
like Peierls transition, superconductivity, electron-phonon
interaction and the interplay between them in a low-dimensional
system. Despite the great interest in these materials progress has
been hindered until recently due to the difficulty to produce
carbon nanotubes with well defined radii as well as due to the
difficulty to determine in detail the structure of nanotubes
present in a given sample. Recently, however, very promising
progress in identifying the structure of nanotubes has been made
by combining Raman and photoluminescence measurements
\cite{Bachilo}. Size selection has been also achieved in certain
cases. For example, growing nanotubes in zeolite crystals has made
it possible to produce tubes with a very narrow radii distribution
\cite{Wang}
thus allowing for a detailed comparison with modern density
functional theory (DFT) based calculations \cite{Liu,macho02,Ye}. Recently
superconductivity has been reported for nanotubes with radii of 4
\AA\ \cite{Tang}. This immediately rises the question as to the
origin of superconductivity, the importance of Peierls distortions
and of electron-electron correlations. Dealing with
electron-lattice and strong electron-electron interaction in a
materials specific way from ab-initio methods has not been
possible so far even for much simpler systems than nanotubes. Thus
in the past either the electron-electron aspect has been
emphasized \cite{Egger}, however, restricted to model studies or
the electron-lattice interaction has been treated in greater
detail dealing with correlations only on the level of local
density approximation (LDA) based functionals \cite{Dubay,Dubay2003}.

In the latter case, however, modern DFT based methods allow for
the parameter free microscopic calculation of phonon modes. So
far, for nanotubes this scheme has been used nearly exclusively
for the determination of Raman active modes. The results are
generally in good agreement with available Raman data thus
emphasizing the reliability of this approach. Calculation of the
full phonon dispersion for a given nanotube is a much more
elaborate task. Based on supercell calculations the phonon
dispersion for a small number of tubes could be determined
recently \cite{Dubay2003,Ye}. This approach, however, has a
certain disadvantage if one is interested in phonon anomalies and
electron-phonon coupling. Anomalies show up in most cases at
phonon wave vectors which are not commensurable with the
underlying lattice, thus even for approximate treatments huge
supercells would be needed to study for example a Peierls
transition by this method. 
Only in very favorable cases, these anomalies appear at high-symmetry
points and can be dealt with accurately by the supercell approach
(as, e.\ g., for graphite \cite{Dubay2003,mault04}).  
Also the calculation of
superconducting properties in the framework of the Eliashberg theory
requires a detailed knowledge of the phonon dispersion over the whole
Brillouin zone (BZ). Especially systems which show phonon anomalies
require usually a very dense mesh of phonon wave vectors which can
not be obtained with supercell methods. The calculations get even
more demanding if the phonon anomalies are related to special nesting
features of the Fermi surface. Since in one-dimensional systems the
Fermi surface consists only of isolated points the most extreme case
for nesting properties is reached here. This requires also a very dense
k-point mesh for calculating the electronic bandstructure and wave
functions. Already for simple systems like graphene and graphite this
effect can be seen easily.  It is reflected in the high sensitivity of
certain phonon modes at the K point to sampling effects of the Fermi
surface as reported recently \cite{Dubay2003,mault04}.

An alternative method is offered by using density functional
perturbation theory (DFPT) which allows for the calculation of
phonon modes at arbitrary wave vectors without relying on large
supercells \cite{Baroni}. Using this approach we have studied in
detail the complete lattice dynamics and electron-phonon
interaction for a (3,3) nanotube. This belongs to the class of
small radii tubes which have been reported to be superconducting.
Our study will try to answer the question whether or not
electron-phonon coupling can explain superconductivity in this
tube. A special complication arises from the Peierls instability
which has to show up in all one-dimensional metallic systems.
However, this has not received much attention in the past since
general believe was that the mean-field Peierls transition
temperature for nanotubes is always very low so that it is of no
practical consequence. These arguments are based on information
from graphite using the folding concept \cite{Saito}. However,
this approach breaks down for tubes with small diameter which is
seen e.g. in the prediction of the wrong ground state for small-radii
nanotubes \cite{Liu,Blase}. Recently an approximate treatment of
the lattice dynamics for (5,0), (6,0) and (5,5) tubes has been
presented which showed for (5,0) tube indeed that the estimated
transition temperature is of the order of 160 K and thus not
negligible \cite{Barnett}. However this approach was based on a
non-selfconsistent tight-binding scheme and did use only an
approximate treatment of the polarization eigenvectors thus
results might be questioned.

In contrast, our DFT-based method allows for a consistent
calculation of electronic states, phonon modes and electron-phonon
coupling without introducing approximations beyond the LDA level.
In this paper we present fully ab-initio results for phonon
dispersion and eigenvectors as well as for the electron-phonon
coupling for the (3,3) nanotubes using a well tested
norm-conserving pseudopotential of Hamann-Schl\"uter-Chiang type
\cite{Holzwarth}. We use DFPT in the mixed-basis pseudopotential
formalism \cite{Heid} which has been successfully applied to study
electron-phonon mediated superconductivity
\cite{Bohnen2001,Bohnen}. As basis functions localized 2s and 2p
functions are used together with plane waves up to an energy
cut-off of 20 Ry. For integration over the BZ a Gaussian broadening
scheme is employed. As test of the reliability of our phonon
approach we have calculated the phonon dispersion for graphene.
Comparison with results published recently show excellent
agreement among the different calculations except for the highest
mode at the K point which is very sensitive to k-point mesh and
broadening as already mentioned \cite{Dubay2003,mault04}. With a very dense
k-point mesh of 5184 points in the BZ we found still a fluctuation
of 1.5 meV for the highest mode when going from a broadening of 0.05
to 0.2 eV. The authors of Ref.~\cite{Dubay2003}  tried to avoid the
complications due to very dense k-point sampling by increasing the
broadening, however, for studying instabilities due to Fermi surface
nesting that is not a practical way since any instability will be
broadened substantially. Thus one of the complication for the
calculations for the nanotube is the requirement of a very dense
k-point mesh and a small broadening.

\begin{figure}
\centerline{\includegraphics[height=1.5in,clip=true]{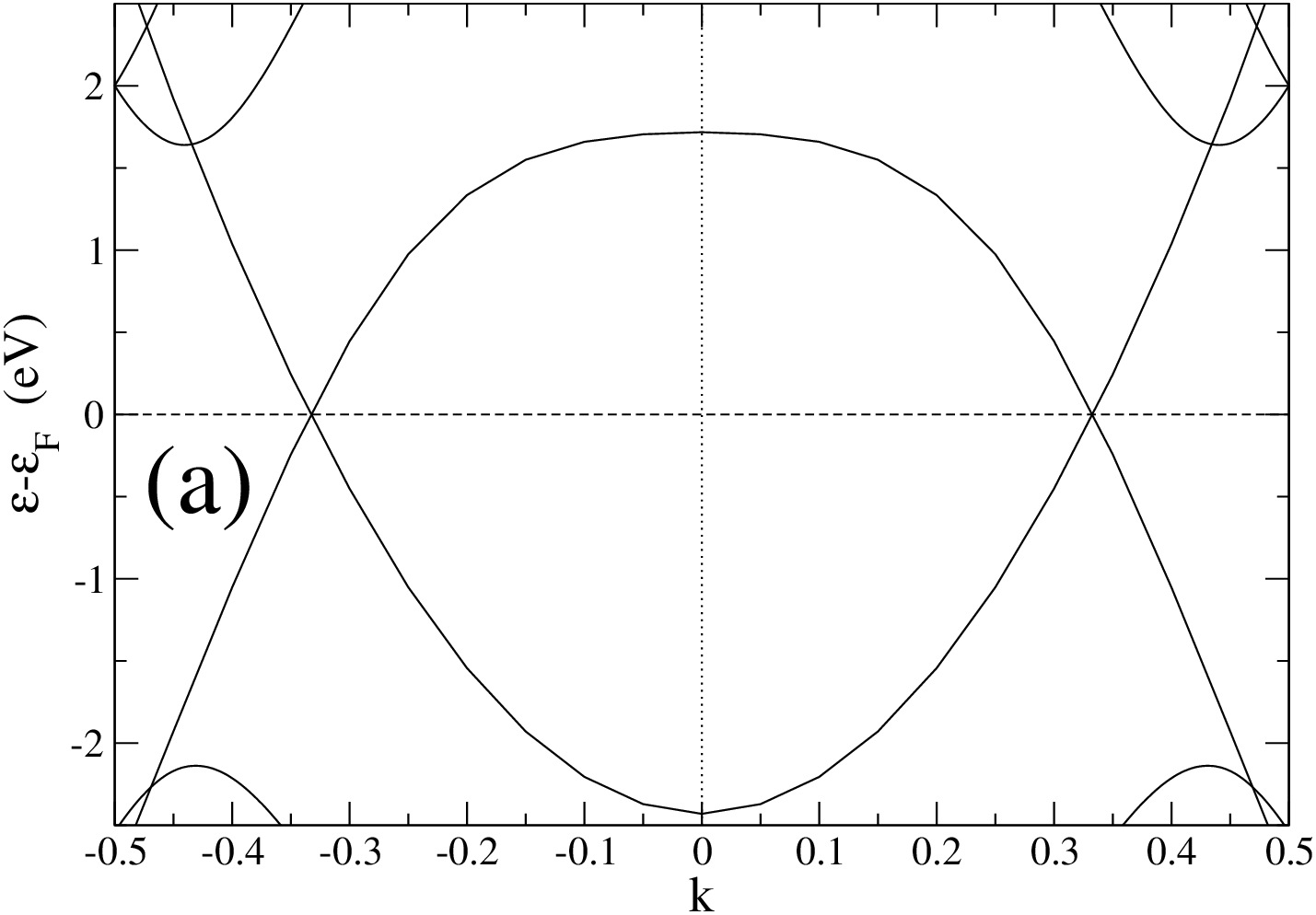}}
\centerline{\includegraphics[height=1.5in,clip=true]{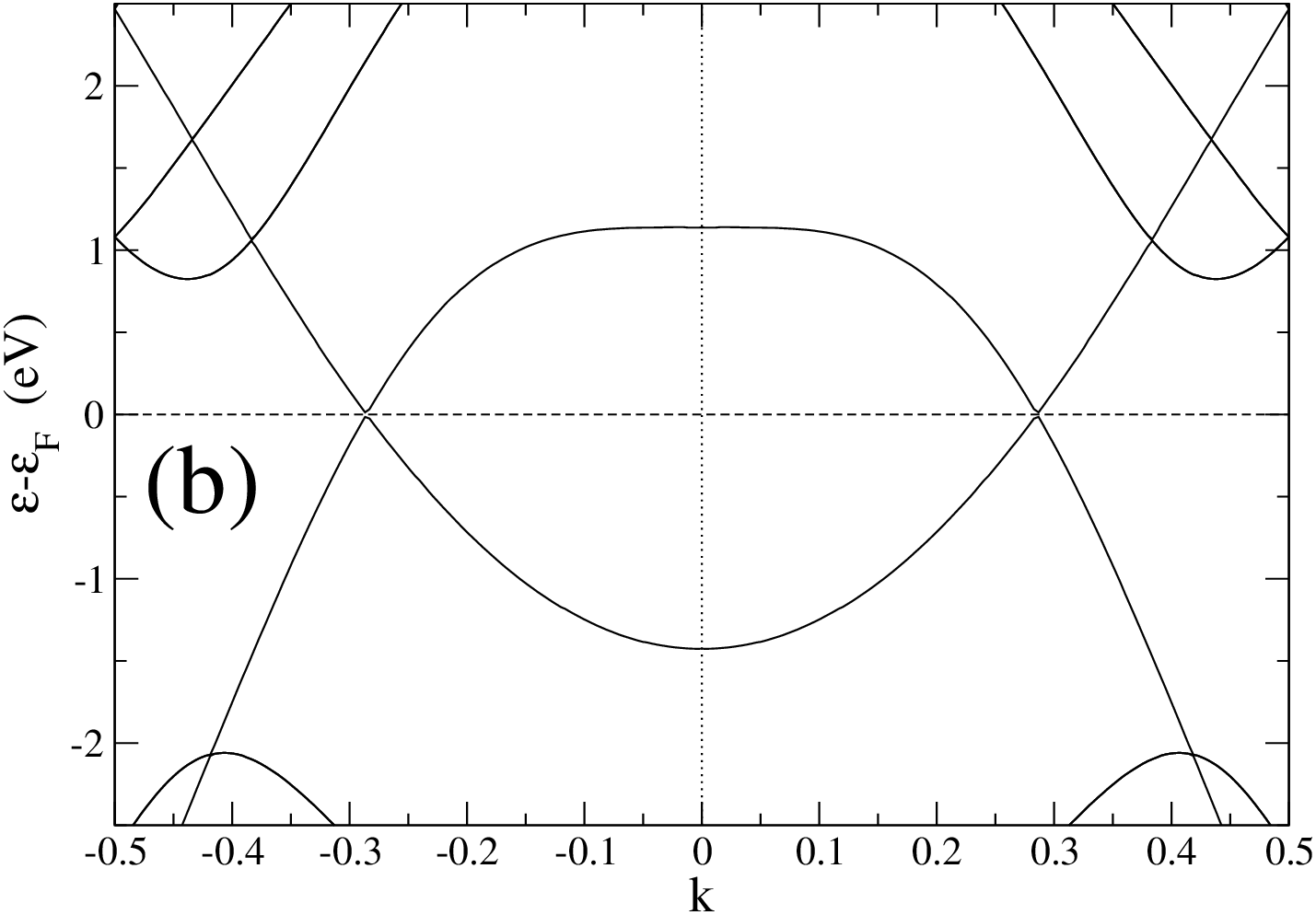}}
\caption{Calculated bandstructure of the (3,3) nanotube. Compared are
results obtained (a) from the calculated
graphene bandstructure using the folding method, and (b) for the true nanotube
geometry.} 
\label{firstfigure}
\end{figure}

Our calculations for the (3,3) nanotubes were done in a supercell
geometry so that all tubes are aligned on a hexagonal array with a
closest distance between adjacent tubes of 10 \AA. The tube-tube
interactions are very small \cite{Liu}. We used 129 k points in
the irreducible part of the BZ and a broadening of 0.2 eV. The
structure was fully relaxed and the optimal geometry agreed very
well with those given in Refs.~\cite{Liu,macho02}. The phonon calculation was
carried out with the DFPT method. For calculation of the
electron-phonon matrix elements we even used up to 1025 k points.
In Fig.~\ref{firstfigure} we have plotted the bandstructure close
to the Fermi energy and for comparison also the bandstructure as 
obtained by using the folding method. 
Our results agree well with those obtained previously \cite{Liu,macho02}.
At k=k$_F$=0.284 (in units
of ${2\pi}/{a}$ where $a$ is the lattice constant of the
graphene honeycomb lattice) we see that two bands are crossing
$\epsilon_F$. This special feature which holds for all (n,n) tubes
is important for phonon anomalies seen at the $\Gamma$ point as
will be emphasized later. The dispersion differs most notably from
those obtained by using the folding technique by a shift of the
k$_F$ value and a change in slope of the two bands crossing at
$\epsilon_F$. This has of course drastic consequences for the
phonon modes.

\begin{figure}
\centerline{\includegraphics[height=1.8in,clip=true]{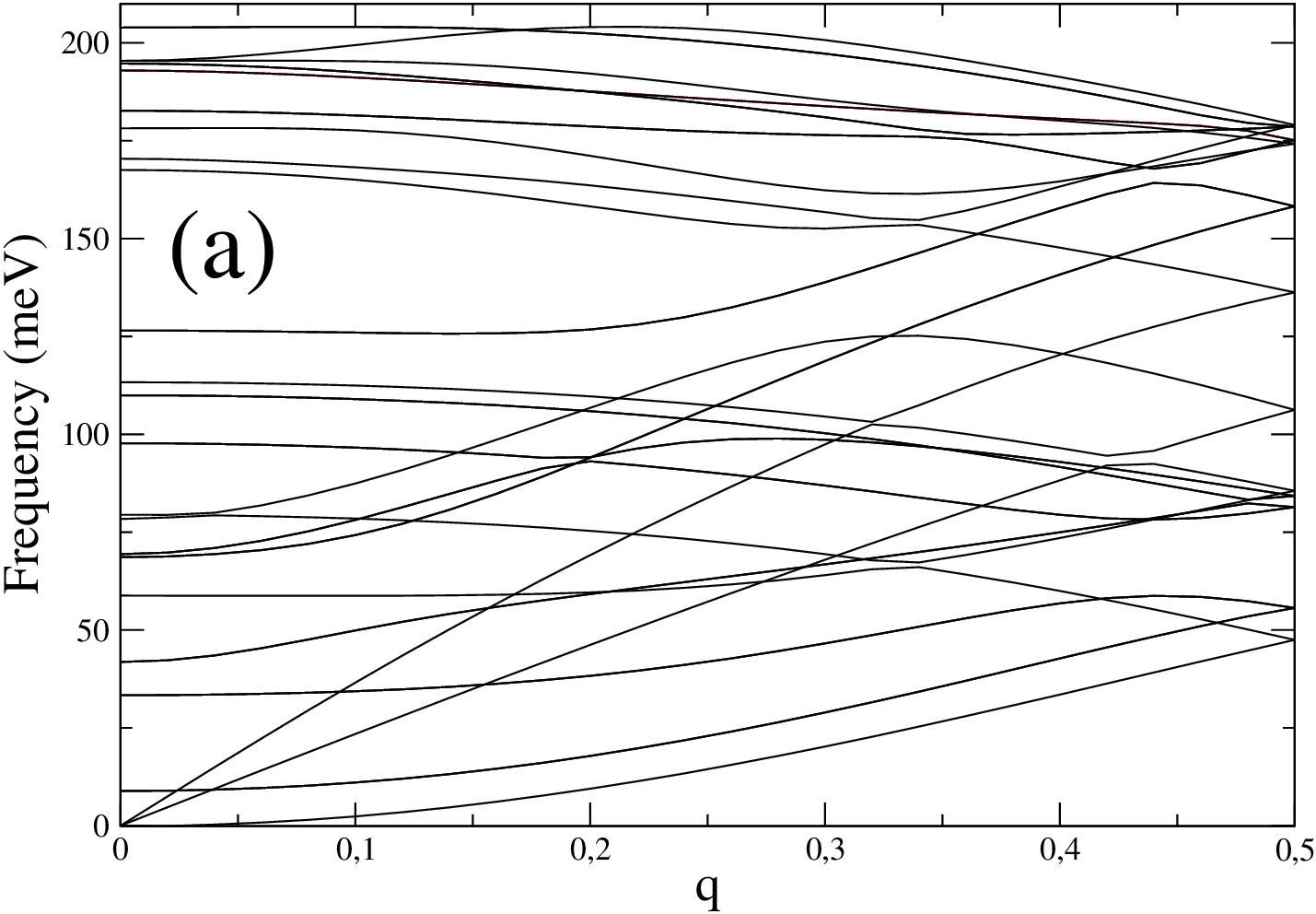}}
\centerline{\includegraphics[height=1.8in,clip=true]{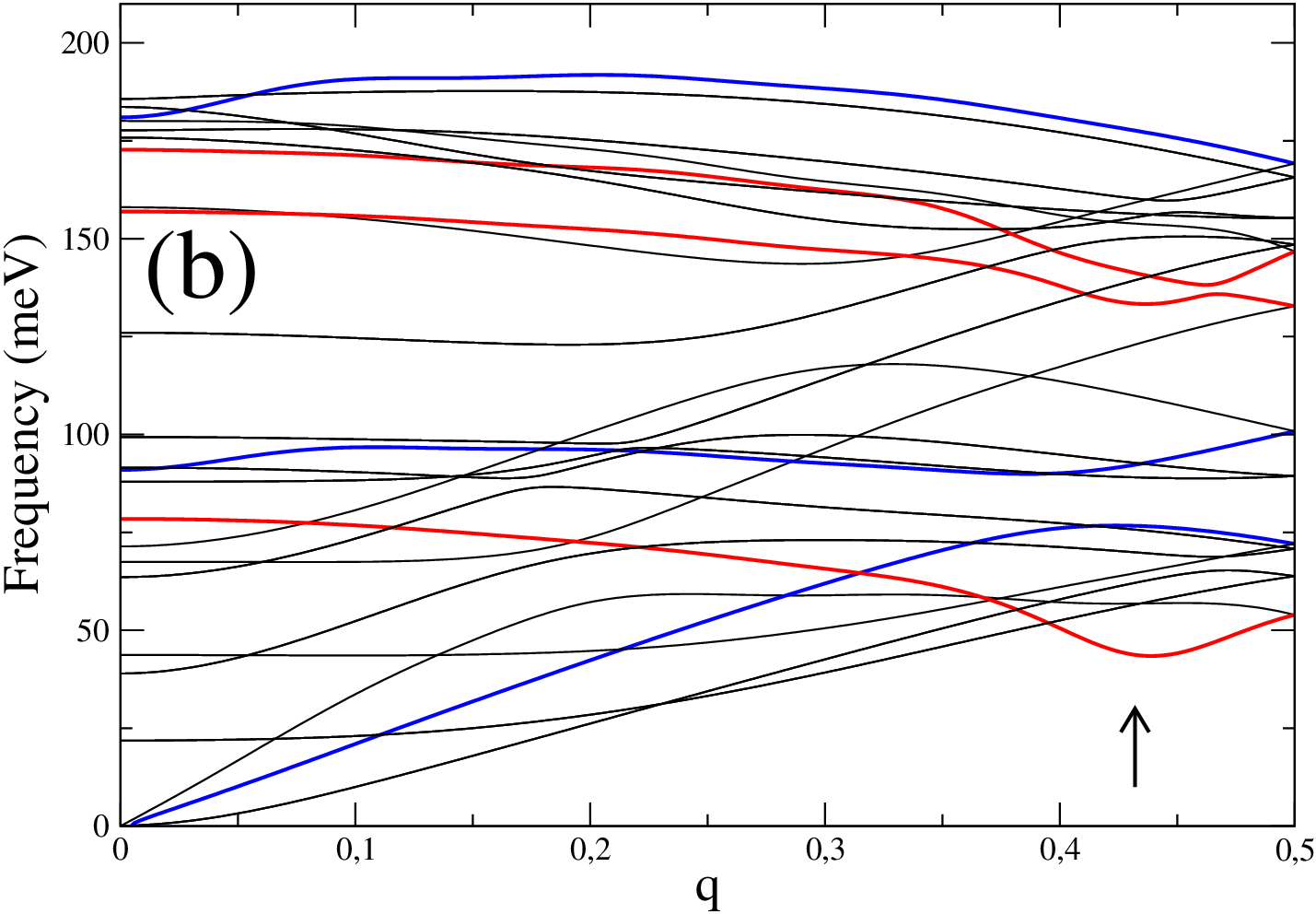}}
\caption{Calculated phonon dispersion curves of the (3,3) nanotube as
obtained (a) by the folding method from the graphene phonon dispersions,
and (b) for the true nanotube geometry.
The latter corresponds to a high effective temperature of 1096\,K.
The arrow indicates 2k$_F$ (folded back to the first Brillouin zone).
} 
\label{secondfigure}
\end{figure}

Phonon results are shown in Fig.~\ref{secondfigure}. Again we have
plotted folding results together with the ab-initio dispersion
curves. These ab-initio results were obtained by Fourier interpolation
on a 16-point grid. The electronic wave functions were obtained from a
calculation with a broadening width of 0.2 eV. In general, the ab-initio
phonon frequencies are softer than the folding results.  Furthermore,
near q=2k$_F$ one sees phonon anomalies in certain phonon branches as
well as a certain softening at the $\Gamma$ point which is not present
in the dispersion obtained by folding. The $\Gamma$-point softening
has been seen and studied already in Ref.~\cite{Dubay,Dubay2003}
for the G band, however, this effect is also present in another mode
at $\approx$90 meV with the same symmetry.  It is related to the fact
that these $\Gamma$-point phonons couple the two electronic bands crossing
right at k$_F$ (Fig.~\ref{firstfigure}), leading to a nesting vector
q=0. Therefore, this softening also indicates an underlying Peierls
instability, however, the corresponding Peierls distortion does not
change the lattice periodicity. Note that in our DFPT approach, the
calculations of phonons include all screening effects due to lattice
distortions in contrast to other treatments like e.g. tight-binding
methods (see Ref.~\cite{Barnett}).

\begin{figure}
\centerline{\includegraphics[height=1.9in,clip=true]{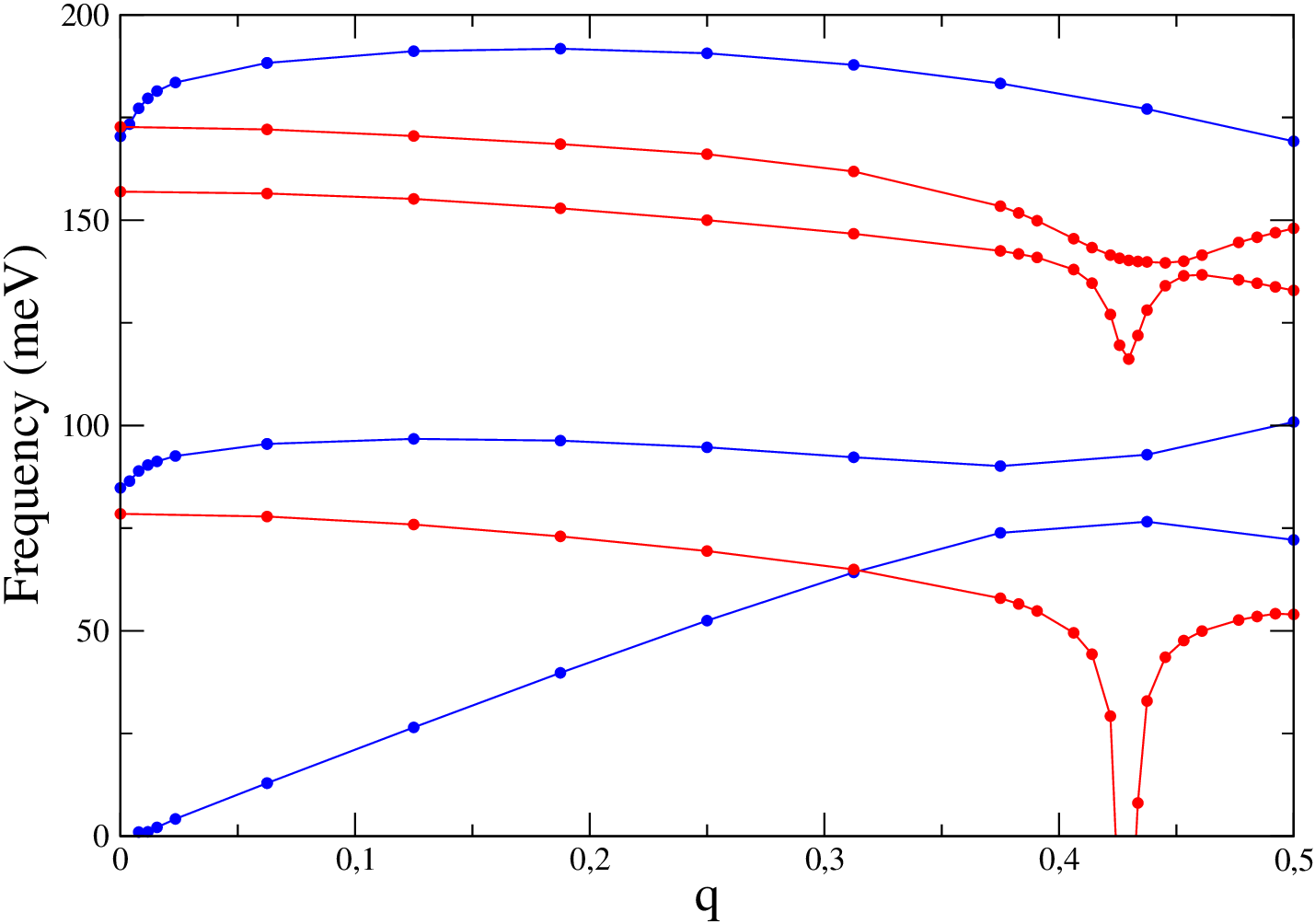}}
\caption{Phonon dispersion curves for the two symmetry classes which
are affected by electron-phonon coupling. Shown are results obtained on a
fine q grid and for a small effective temperature of 137\,K.} 
\label{thirdfigure}
\end{figure}

To study the anomalies near q=0 and q=2k$_F$ in more detail we
have increased the number of k points and reduced the broadening 
from 0.2 to 0.025 eV. This can be interpreted as a variation of the
electronic temperature from 1096 to 137\,K. With finer sampling and
reduced temperature
one of the modes with the anomalous behavior at
q=2k$_F$ gets unstable. Results for T$_{el}$=137\,K are shown in
Fig.~\ref{thirdfigure}. Here only those modes which are sensitive
to the temperature variation are shown. The modes which show
anomalous behavior at the $\Gamma$ point stay stable for all
temperatures studied but should eventually go soft at low enough
temperatures. Using the definition of the Peierls transition
temperature (T$_P$) with $\omega_{2k_F}$(T$_P$)=0 leads to a lower
limit of 137 K for T$_P$ which is of similar magnitude as the
value given for a (5,0) tube in Ref.~\cite{Barnett}. It is important to note
that, contrary to early estimates of T$_P$ for nanotubes of the order of
1\,K \cite{mintm92}, this temperature is of sizeable magnitude and thus the
Peierls instability might compete with the superconducting
transition.

To investigate the possibility of phonon-induced superconductivity,
we have calculated the microscopic electron-phonon
coupling parameters, which are central to the Eliashberg theory of
strong-coupling superconductivity. 
The coupling constant for a phonon mode q$\lambda$ is given by
\begin{equation}
\lambda_{q\lambda} = \frac {2}{\hbar
N(\epsilon_F)\omega_{q\lambda}} \sum_{k\nu\nu'} |g^{q\lambda}_{
k+q\nu'; k\nu}|^2 \delta(\epsilon_{k\nu})
\delta(\epsilon_{k+q\nu'}),
\label{eq1}
\end{equation}
where N($\epsilon_F$) is the density of states at the Fermi energy
and all energies are measured with respect to $\epsilon_F$. The
electron-phonon coupling matrix element is given by
\begin{displaymath}
g^{q\lambda}_{k+q\nu'; k\nu}\approx<\phi_{k+q \nu'}| \delta
V^{q\lambda}_{eff} \mid \phi_{k\nu}>
\end{displaymath}
with $\delta V^{q\lambda}_{eff}$ being the change of the
effective crystal potential due to a phonon q$\lambda$ and
$\mid\phi_{k\nu}>$ being the electronic wave functions . Since this
quantity is very sensitive to k-point sampling we have used a grid
of 1025 points. Due to the selfconsistent determination of
$\delta$V, the matrix elements include all screening effects in
contrast to tight-binding approaches. 
Because the Fermi surface consists only of the points k=$\pm$k$_F$, 
contributions to the total electron-phonon coupling constant
$
\lambda=1/{N_q} \sum_{q\lambda} \lambda_{q\lambda}
$
are restricted to q=0 and q=2k$_F$.
Roughly 20\% comes from q=0 while 80\% is contributed from 2k$_F$. 
An estimate for T$_c$ can be obtained by using the Allen-Dynes
formula \cite{Allen}.
Neglecting the effect of the electron-electron interaction on the
pairing, it gives an upper limit of T$_c$=0.833 $\omega_{\ln}
\exp \{- 1.04 (1+1/\lambda)\}$, where
the effective phonon frequency is defined as
\begin{displaymath}
\omega_{\ln} = \exp \{\frac{1}{\lambda}
\frac{1}{N_q}\sum\limits_{q\lambda} \lambda_{q\lambda} \ln
(\omega_{q\lambda})\}   .
\end{displaymath}
Using our results for T$_{el}$=1096\,K gives $\lambda\approx 0.25$ and
$\omega_{\ln}\approx 60.4$ meV and as an upper limit T$_c=3$\,K.
One has, however, to consider the possibility that the presence of the Peierls
instability significantly alters this estimate.
This is because in the expression (\ref{eq1}) for the coupling constant
$\lambda_{q\lambda}$ the renormalized phonon frequencies enter.
The softening of the critical mode thus 
produces a strong temperature dependence of $\lambda$, which formally
diverges at T$_P$. At the same time $\omega_{\ln}$
goes to zero.
Simulation of this mode softening effect using the Allen-Dynes formula
shows that T$_c$ can be enhanced due to the softmode, but the
obtained maximal value of $\approx 30$\ K is still significantly
smaller that T$_P$.
Taking into account a finite electron-electron interaction further reduces
this value.

To get a more precise estimate of the Peierls transition temperature we have
made connection with simple model studies which have been carried
out in the past considering a Fr\"ohlich Hamiltonian in random-phase 
approximation \cite{last}. 
Extending this approach to the present 2-band case gives for the
the critical mode at q=2k$_F$ a temperature
dependence of the frequency as
\begin{displaymath}
\{\hbar\omega_{q=2k_F}(T)\}^2 = A\ln(T/T_P)\,,
\end{displaymath}
where
\begin{displaymath}
A=2 \hbar \omega^{bare} \sum\nolimits_\nu (g_\nu)^2 N_\nu(\epsilon_F)\,.
\end{displaymath}
Here, N$_\nu(\epsilon_F)$ is the partial density of states per spin of band
$\nu$, $\omega^{bare}$ is  
the unrenormalized phonon frequency,
and $g_\nu$ is the electron-phonon matrix element for
scattering processes from k$_F$ to -k$_F$ within the $\nu$-th band
induced by the critical phonon at q=2k$_F$. Due to symmetry, 
only intra-band scattering processes contribute.
Estimates of the parameters from our ab-initio calculations for different
effective temperatures lead to T$_P\approx$240 K and A$\approx$(39
meV)$^2$.
Note that the expression for A does not depend on the phonon frequency
but only on the phonon eigenvector due to the fact that $(g_\nu)^2\sim
1/\omega^{bare}$.
Our finding that T$_P$ is an order of magnitude larger than T$_c$
suggests that for the (3,3) nanotube the Peierls instability
dominates over the superconducting one.
%

In summary we have presented here a fully ab-initio calculation of
the lattice dynamics and the electron-phonon coupling for the
(3,3) carbon nanotube. Without relaying on simplifying
approximations the Peierls transition could be seen and a Peierls
mean field transition temperature T$_P\approx 240$ K was
predicted. Calculating the electron-phonon coupling using the same
scheme resulted in a superconducting transition temperature which was
one order of magnitude smaller than T$_P$.
This makes it not very likely that
superconductivity based on the electron-phonon mechanism is
present in (3,3) nanotubes.
This does, however, not exclude the possibility that phonon-mediated
superconductivity exists in the case of the (5,0) nanotube, which is
another member of the class of a 4 \AA\ tubes for which superconductivity
has been observed experimentally \cite{Tang}.
Effects which have been discussed here are not
accessible with the folding method since they depend sensitively
on the curvature of the nanotubes, the detailed bandstructure and
nesting features. Since in our approach arbitrary q points are
treated on the same footing we could easily see that the q=0
anomaly is only a special case of strong electron-phonon coupling.
In contrast to former studies we could show that this effect is
not restricted to the G band but should be seen in another mode,
too, belonging to the same symmetry class.

CTC and LHJ are supported by RGC Hong Kong through HKUST6152/01P.

\end{document}